\documentclass[aps,prl,preprint,groupedaddress]{revtex4}
\usepackage{graphicx}
\begin{document}

%\preprint{}

\title{Two-point resistance of a cobweb network with a $2r$ boundary}


\author{ Zhi-Zhong Tan  \footnote{E-mail: tanz@ntu.edu.cn ; ~~ tanzzh@163.com}}
\address{ Department of Physics, Nantong University, Nantong 226019, China }

\date{\today}

\begin{abstract}
We consider the problem of two-point resistance on an $m\times n$ cobweb network with a $2r$ boundary which has never been solved before. Past efforts prior to 2014 researchers just only solved the cases with free boundary or null resistor boundary. This paper gives the general formulae of the resistance between any two nodes in both finite and infinite cases using a method of direct summation pioneered by Tan[Z. Z. Tan, et al, J. Phys. A 46, 195202 (2013)], which is simpler and can be easier to use in practice. This method contrasts the Green¡¯s function technique and the Laplacian matrix approach, which is difficult to apply to the geometry of a cobweb with a 2r boundary. We deduced several interesting results according to our general formula. In the end we compare and illuminate our formulae with two examples. Our analysis gives the result directly as a single summation, and the result is mainly composed of the characteristic roots.
\\
\\
\noindent{\bf Key words:}  two-point resistance, cobweb, matrix equation, direct method,  boundary conditions..
\\
\noindent{\bf PACS numbers:}  05.50.+q,  84.30.Bv,  01.55.+b,  02.10.Yn

\end{abstract}

\maketitle


\section{1. Introduction}
A classic problem in electric circuit theory studied by numerous authors for more than  160 years is the computation of the resistance between two nodes in a resistor network[1-21]. Today the research on the resistor network is no longer confined to the circuit field and it has been expanded into the basic model in various disciplines[3-21]. Modeling resistor network can help to carry on the research of exact finite-size corrections in critical systems[17-21], the researches of graphene network, the researches of the structure of some metal compound crystals or non-metallic crystals, the research of the structure of the multiferroic magnetoelectric material, the structure of fullerenes $(C_{60})$, the researches of carbon nano-tube[7] and so on. However, it is usually very difficult to find  the exact resistance formula of a resistor networks as it is an interdisciplinary problem[3-12].

Past efforts prior to 2004 were focused mainly on infinite lattices and the use of Green¡¯s function technique[1,2,14-16]. Little attention has been paid to finite networks[3-12], even though the latter are those occurring in real life. It was not until 2004 that Wu[3] revisited the resistance problem of two arbitrary nodes, and established a theorem to compute the equivalent resistance of the resistor network with normative boundary. Using the theorem to compute the equivalent resistance relies on two Laplacian matrix along orthogonal direction, and the resistance is expressed by double summation [3-6,10]. Especially it is generally difficult to solve the eigenvalue problem for non-regular networks such as a cobweb with an arbitrary boundary[12].

In order to solve this dilemma, we[7-9] build a new independent method to calculate the two-point resistance, using the method to compute the equivalent resistance relies on just one matrix along one direction, and the resistance is expressed by single summation. Especially, when just one boundary or a pair of boundary ( opposite ) condition changes, our method is still valid to computation of the resistance. For example, in 2013 a cobweb network was proposed by one of us[7-9], although the problem of the cobweb network has been solved by the Wu's method [6] by means of double summation, however, soon afterwards ref.[11] solved the cobweb problem  by the direct method of single summation.  Recently, a globe network is solved by the direct method of single summation [12]. In this paper we apply the direct method to the cobweb with $2r$ boundary  which has never been solved before.

We first research a general matrix $A_{m}$  by summarizing the direct method as we did before[7-9,12]. Consider  an $m\times n$ cobweb resistor network which has $n$ radial line and $m$ polygons. Bonds in the radial and arc directions represent, respectively,  resistors $r_{0}$ and $r$ except for boundary ( the boundary resistor is $r_{1}$  ) as shown in fig.1. According to Kirchhoff's law ( $KCL$ and $KVL$ ) to model the difference equations of the electric currents along the radial line direction, we  therefore obtain the key matrix
\begin{eqnarray}
{\bf A}_{ m}=
\left( {\begin{array}{cccccc}
   {{2+h}} & {{-h}} &{{0}}&{{0}}& {\cdots}&{{0}} \\
   {{-h}} & {{2+2h}} & {{-h}} &{{0}} &{\cdots} & {\cdots} \\
   {\vdots} & {\vdots} &{\ddots} &{\ddots} & {\vdots} & {\vdots}\\
   {{0}} & {\cdots} & {{0}} &{{-h}} & {{2+2h}} & {{-h}} \\
   {{0}} & {\cdots} & {{0}} &{{0}} & {{-h}} & {{2+h+h_{1}}} \\
\end{array}} \right),
\end{eqnarray}
where $r/r_{0}=h $, $r_{1}/r_{0}=h_{1} $. In fig.1 $r_{1}$ is arranged on the boundary .

It is must to conduct the diagonalization of matrix $A_{m}$  and to find the explicit eigenvectors and eigenvalues. Obtaining the explicit eigenvectors is the key to the problem. We find several cases is easy to solve the problem when $h_{1}=\{0,1,2\}$, but the problem  has been researched  respectively by [8-12] except for $h_{1}=2 $. In this paper we will study the cases of $h_{1}=2$  to apply to the cobweb problem.

The organization of this paper is as follows: In section 2 , we present a general exact formula of the resistance between any two nodes in the cobweb network with a $2r$ boundary. In section 3, we prove the equivalent resistance by establishing matrix equation and matrix transform methods, and illustrate the formula by two examples. Section 4 gives a summary and discussion of the method and results.

\begin{figure*}
\begin{center}
\includegraphics[width=6cm,bb=0 0 183 184]{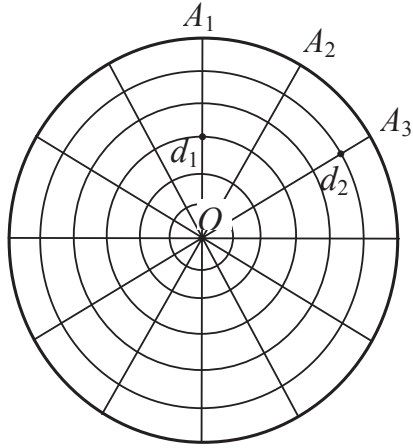}
\caption{ An $6\times 12$ cobweb network with a $2r$ boundary, which has $12$ radial line and $6$ circle (including boundary). Bonds in the radial and arc directions represent, respectively,  resistors $r_{0}$ and $r$ except for a $2r$ boundary . }
\end{center}
\end{figure*}

\section{2. The general resistance formulae }
Consider an $m\times n$ cobweb resistor network with a $2r$ boundary, which has $n$ radial line and $m$ polygons  (or circle ). Bonds in the radial and arc directions represent, respectively, resistors $r_{0}$ and $r$ except for a $2r$ boundary, and let $O$ be the origin of the coordinate system as shown in fig.1. We find the resistance between any two nodes $d_{1}(0,y_{1}) $ and $ d_{2}(x,y_{2})$, where $\{x, y\}$ are coordinates, can be written as
\begin{eqnarray}
R_{m \times n} \left( {\{ 0,y_1 \} ,\{ x,y_2 \} } \right) = \frac{{2r}}{m}\sum\limits_{i = 1}^m {\frac{{F_n^{(i)} (S_{1,i}^2  + S_{1,i}^2 ) - 2(F_x^{(i)}  + F_{n - x}^{(i)} )S_{1,i} S_{2,i} }}{{\lambda _i^n  + \bar \lambda _i^n  - 2}}} ,
\end{eqnarray}
where  $ h={r}/{r_{0}}$, $\lambda_{i}\cdot\bar{\lambda}_{i}=1 $,  and
\begin{eqnarray}
F_k^{(i)}  = {{(\lambda _i^k  - \bar \lambda _i^k )} \mathord{\left/
 {\vphantom {{(\lambda _i^k  - \bar \lambda _i^k )} {(\lambda _i  - \bar \lambda _i )}}} \right.
 \kern-\nulldelimiterspace} {(\lambda _i  - \bar \lambda _i )}}, \qquad
S_{k,i}  = \sin (y_k \theta _i ),\\
\theta _i  = {{(2i - 1)\pi } \mathord{\left/
 {\vphantom {{(2i - 1)\pi } {(2m)}}} \right.
 \kern-\nulldelimiterspace} {(2m)}}
\qquad i=1,2.3,...m. \qquad \nonumber\\
\lambda_{i}=1+h-h\cos\theta_{i}+\sqrt{(1+h-h\cos\theta_{i})^{2}-1} .
\end{eqnarray}
When $y_{1}, y_{2}$ and $x$ are special coordinates, we have the special cases:

{\bf Case 1}. when $n\rightarrow \infty$  with $x, y$  finite, we have
\begin{eqnarray}
R_{\infty  \times n} \left( {\{ 0,y_1 \} ,\{ x,y_2 \} } \right) = \frac{r}{m}\sum\limits_{i = 1}^m {\frac{{S_{1,i}^2  + S_{1,i}^2  - 2\bar \lambda _i^x S_{1,i} S_{2,i} }}{{\sqrt {(1 + h - h\cos \theta _i )^2  - 1} }}} .
\end{eqnarray}

{\bf Case 2}. when $m, n\rightarrow \infty$, but $x$  and $y_{1}-y_{2}$ are  finite,  we have
\begin{eqnarray}
R_{\infty  \times \infty } (\{ 0,y_1 \} ,\{ x,y_2 \} ) = \frac{r}{\pi }\int_0^\pi  {\frac{{1 - \bar \lambda _\theta ^x \cos (y_1  - y_2 )\theta }}{{\sqrt {(1 + h - h\cos \theta )^2  - 1} }}} d\theta .
\end{eqnarray}
where
$\bar \lambda _\theta   = 1 + h - h\cos \theta  - \sqrt {(1 + h - h\cos \theta )^2  - 1} . $

{\bf Case 3}. When $d_{1}$ and $d_{2}$ are on the same radial at $(0,y_{1}) $ and $ (0,y_{2})$, we have
\begin{eqnarray}
R_{m \times n} (\{ 0,y_1 \} ,\{ 0,y_2 \} ) = \frac{{2r}}{m}\sum\limits_{i = 1}^m {\left( {\sin y_1 \theta _i  - \sin y_2 \theta _i } \right)^2 \frac{{\coth (n\ln \sqrt {\lambda _i } )}}{{\lambda _i  - \bar \lambda _i }}} .
\end{eqnarray}

{\bf Case 4}. when the nodes $d_{1}$ and $d_{2}$  are respectively at the center and the edge, we have
\begin{eqnarray}
R_{m\times n} (O,A_{m})=\frac{2r}{m}\sum_{i=1}^{m}\frac{\coth(n\ln\sqrt{\lambda_{i}})}{\lambda_{i}-\bar{\lambda}_{i}}.
\end{eqnarray}

{\bf Case 5}. When $d_{1}$ and $d_{2}$  are on the same arc line at $(0, y) $ and $ (x, y)$, we have
\begin{eqnarray}
R_{m \times n}^{{\rm{arc}}} (\{ 0,y\} ,\{ x,y\} ) = \frac{{4r}}{m}\sum\limits_{i = 1}^m {\bigg( {\frac{{F_n^{(i)}  - (F_x^{(i)}  + F_{n - x}^{(i)} )}}{{\lambda _i^n  + \bar \lambda _i^n  - 2}}} \bigg)} \sin ^2 (y\theta _i ) .
\end{eqnarray}

{\bf Case 6}. when the nodes  $d_{1}$ and $d_{2}$  are at the edge, we have
\begin{eqnarray}
R_{m \times n}^{{\rm{edge}}} (\{ 0,m\} ,\{ x,m\} ) = \frac{{4r}}{m}\sum\limits_{i = 1}^m {\frac{{F_n^{(i)}  - (F_x^{(i)}  + F_{n - x}^{(i)} )}}{{\lambda _i^n  + \bar \lambda _i^n  - 2}}} .
\end{eqnarray}

\section{3. Derivation of the resistance formula }

\subsection{3.1  Designing the virtual currents }
We assume the electric current $J$ is constant and goes from the input $d_{1}(0,y_{1})$ to the output $d_{2}(x,y_{2})$ as shown in fig.1. Denote the currents in all segments of the network as shown in Fig.2.  The currents passing through all resistor $r$ in the sides of the $m$ polygons ( from center to edge ) are respectively: $ I_{A_{1},k} , I_{A_{2},k} , \cdots , I_{A_{m},k},(1\leq k\leq n) $ ; the currents passing through the resistors $r_{0}$ of the $n$ radial lines are respectively: $ I_{k}^{(1)}, I_{k}^{(2)}, \cdots, I_{k}^{(m)}, (0\leq k\leq n) $.

To find the resistance $R_{m\times n}(d_{1}, d_{2})$ we use the indirect method calculating the voltages to  realize it.  The voltages between $d_{1}$, $d_{2}$, and the center $O $  are, respectively,
$$ U_{m\times n}(O, d_{1})= r_{0}\sum_{i=1}^{y_{1}}I_{0}^{(i)}, \qquad U_{m\times n}(O, d_{2})=r_{0}\sum_{i=1}^{y_{2}}I_{x}^{(i)}.   $$
where $I_0^{(i)}$ and $I_{x}^{(i)}$ denotes currents along the  radial, respectively, via $d_{1}$ and  $d_{2}$.
It then follows from the Ohm's law that  the resistance between $d_1(0, y_{1})$ and $d_2(x, y_{2})$ is
\begin{eqnarray}
R_{m\times n}(d_{1}, d_{2})=\frac{r_{0}}{J}[\sum_{i=1}^{y_{2}}I_{x}^{(i)}-\sum_{i=1}^{y_{1}}I_{0}^{(i)}].
\end{eqnarray}
How to solve the current parameters  $I_{0}^{(i)}$  and  $I_{x}^{(i)}$   is the key to the problem.  We are going to solve the problem by constructing matrix equation model in terms of Kirchhoff¡¯s law .

\begin{figure*}
\begin{center}
\includegraphics[width=9cm,bb=0 0 229 153]{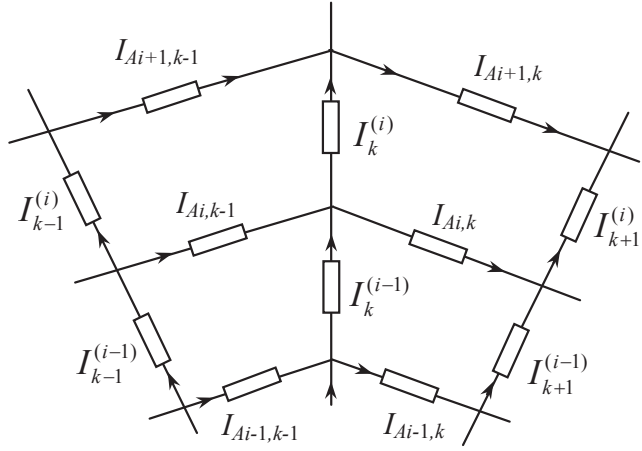}
\caption{ The sub-network model containing the currents direction .}
\end{center}
\end{figure*}

\subsection{3.2 The Matrix Equation and General Solution }

We assume the center $(O)$  is the origin of the coordinate system, and the boundary resistor is  $2r$. Using Kirchhoff's laws ( $KCL$ and $KVL$ ) to study the resistor network, the nodes current equations and the meshes voltage equations can be obtained from Figure 2. We focus on the four rectangular meshes and nine nodes, this gives the relation
\begin{eqnarray}
I_{k + 1}^{(1)}  = (2 + h)I_k^{(1)}  - hI_k^{(2)}  - I_{k - 1}^{(1)}, \qquad (i=1)\qquad  \qquad \nonumber\\
I_{k + 1}^{(i)}  = (2 + 2h)I_k^{(i)}  - hI_k^{(i - 1)}  - hI_k^{(i + 1)}  - I_{k - 1}^{(i)}, \qquad (1<i<m) \nonumber\\
I_{k + 1}^{(m)}  = (2 + 2h)I_k^{(m)}  - hI_k^{(m - 1)}  - I_{k - 1}^{(m)} , \qquad (i=m) .\qquad
\end{eqnarray}
Equations (12) can be written in a matrix form
\begin{eqnarray}
{\bf I}_{k+1}={\bf A}_{ m}{\bf I}_{k}-{\bf I}_{k-1},
\end{eqnarray}
where the bound current is not considered in (13) ( two bound currents are given  respectively by (26) and (27) ), and {\bf I}$_{k} $ denotes a $m\times 1 $  column matrix :
\begin{eqnarray}
{\bf I}_{k}=[I_{k}^{(1)}, I_{k}^{(2)}, I_{k}^{(3)} ,\cdots,I_{k}^{(m)}]^{T},
\end{eqnarray}
and ${\bf A}_{m}$ is an $m\times m$ tridiagonal matrix, from  (12) or (1) it can be written as
\begin{eqnarray}
{\bf A}_{ m}=
\left( {\begin{array}{cccccc}
   {{2+h}} & {{-h}} &{{0}}& {{0}}& {\cdots}&{{0}} \\
   {{-h}} & {{2+2h}} & {{-h}}  &{{0}}& {\cdots} & {\cdots} \\
   {\vdots}&{\ddots} &{\ddots} &{\ddots}&{\ddots}&{\vdots}\\
   {{0}} & {\cdots} &{{0}}& {{-h}} & {{2+2h}} & {{-h}} \\
   {{0}} & {\cdots} &{{0}}& {{0}} & {{-h}} & {{2+3h}} \\
\end{array}} \right),
\end{eqnarray}
\\
where $r/r_{0}=h $. Eq.(13) (14) (15) form the matrix equation model of the $m\times n$ cobweb network with a $2r$ boundary.

Next, we consider the solution of (13) by conducting the matrix transformation by means of the method made in [7,12]. multiplying (13) from the  left-hand side by an $m\times m$ undetermined square matrix ${\bf P}_{m}$ . Thus
\begin{eqnarray}
{\bf P}_{m}{\bf I}_{k+1}={\bf P}_{m}{\bf A}_{m}{\bf I}_{k}-{\bf P}_{m}{\bf I}_{k-1},
\end{eqnarray}
Since ${\bf A}_{m}$  is Hermitian that matrix ${\bf P}_{m}$ can be determined such that
\begin{eqnarray}
{\bf P}_{m}{\bf A}_{m}=diag(t_{1},t_{2},\cdots,t_{m}){\bf P}_{m},
\end{eqnarray}
where  $t_{i}$ is a eigenvalue of matrix ${\bf A}_{m}$.
Solving (17) and obtain
\begin{eqnarray}
t_{i}=2(1+h)-2h\cos\theta_{i} \ , \quad (i=1,2.3\cdots , m),
\end{eqnarray}
\begin{eqnarray}
 {\bf P}_{m}=
\left({\begin{array}{cccc}
   {\cos(1-\frac{1}{2})\theta_{1}} & {\cos(2-\frac{1}{2})\theta_{1}} & {\cdots} & {\cos(m-\frac{1}{2})\theta_{1}} \\
   {\cos(1-\frac{1}{2})\theta_{2}} & {\cos(2-\frac{1}{2})\theta_{2}} & {\cdots} & {\cos(m-\frac{1}{2})\theta_{2}} \\
   {\vdots} & {\vdots} & {\ddots}  &{\vdots} \\
   {\cos(1-\frac{1}{2})\theta_{m}} & {\cos(2-\frac{1}{2})\theta_{m}} & {\cdots} & {\cos(m-\frac{1}{2})\theta_{m}} \\
\end{array}} \right),
\end{eqnarray}
where $ \theta_{i}=\frac{(2i-1)\pi}{2m} $, $i=1,2,3,\cdots,m$ .
A simple calculation shows that the new matrix {\bf P}$_{m}$ is invertible, with the following inverse matrix
\begin{eqnarray}
{\bf P}_{m}^{-1}=\frac{2}{m}
\left({\begin{array}{cccc}
   {\cos(1-\frac{1}{2})\theta_{1}} & {\cos(1-\frac{1}{2})\theta_{2}} &{\cdots} & {\cos(1-\frac{1}{2})\theta_{m}} \\
   {\cos(2-\frac{1}{2})\theta_{1}} & {\cos(2-\frac{1}{2})\theta_{2}} & {\cdots}  & {\cos(2-\frac{1}{2})\theta_{m}} \\
   {\vdots} & {\vdots} & {\ddots}   &{\vdots} \\
   {\cos(m-\frac{1}{2})\theta_{1}} & {\cos(m-\frac{1}{2})\theta_{2}} & {\cdots} & {\cos(m-\frac{1}{2})\theta_{m}} \\
\end{array}} \right) .
\end{eqnarray}
By (16) and (17) we define
\begin{eqnarray}
{\bf{P}}_m {\bf{I}}_k^{}  = {\bf{X}}_k , \quad \text{or}    \quad
{\bf{I}}_k  = {\bf{P}}_m^{ - 1} {\bf{X}}_k .
\end{eqnarray}
After making use of (16) and (17),  we obtain the equation
\begin{eqnarray}
{\bf X}_{k+1}=diag(t_{1},t_{2},\cdots,t_{m}){\bf X}_{k}-{\bf X}_{k-1} ,
\end{eqnarray}
Making use of (18) and (22) the roots of the characteristic equation for $X_{k}$ are solved by
\begin{equation}
\left.
  \begin{array}{c}
  \lambda_{i}=1+h-h\cos\theta_{i}+\sqrt{(1+h-h\cos\theta_{i})^2-1} \hbox{ }\\
  \bar{\lambda_{i}}=1+h-h\cos\theta_{i}-\sqrt{(1+h-h\cos\theta_{i})^2-1}\hbox{ }
  \end{array}
\right\} .
\end{equation}

Next, we consider the solution of (22) in the cases of which inject current $J$ at $d_{1}(0,y_{1})$  and exit the current at $d_{2}(x,y_{2})$ . We therefore need to consider the piecewise solution of (22) and obtain
\begin{eqnarray}
X_{k}^{(i)}&=& X_{1}^{(i)}F_{k}^{(i)} -X_{0}^{(i)}F_{k-1}^{(i)}, ~   ~ 0\leq k\leq x \\
X_{k}^{(i)}&=& X_{x+1}^{(i)}F_{k-x}^{(i)} -X_{x}^{(i)}F_{k-x-1}^{(i)}, ~  ~ x\leq k\leq n
\end{eqnarray}
where $F_{k}^{(i)}=(\lambda_{i}^{k}-\bar{\lambda}_{i}^{k})/(\lambda_{i}-\bar{\lambda}_{i})$ is defined in (3).

\subsection{3.3 Bound conditions with the input and output currents}
While either (13) or (22) serves to determine $I_{k}$ when there is no external current injected to the network, to compute the resistance between nodes $d_{1} = d_{1}(0, y_{1})$ and $d_{2} = d_{2}(x , y_{2})$ we need to inject current $J$ at $d_{1}$ and exit the current at $d_{2}$. Then we have
\begin{eqnarray}
{\bf I}_{n-1}+{\bf I}_{1}&=&{\bf A}_{m}{\bf I}_{0}-J{\bf H_{1}},\\
{\bf I}_{x+1}+{\bf I}_{x-1}&=&{\bf A}_{m}{\bf I}_{x}-J{\bf H_{2}},
\end{eqnarray}
where matrix ${\bf A}_{m}$ is given by (15), and  ${\bf H_{1}, H_{2}}$ are two $m\times 1$  column matrices which can be expressed by
\begin{eqnarray}
{\bf H_{1}}=[\overbrace{0, 0, \cdots0,-h,h}^{\mathrm{from} ~  1\mathrm{th} ~ \mathrm{to} ~ (y_{1}+1)\mathrm{th}},0,\cdots,0]^{T}, \\
{\bf H_{2}}=[\overbrace{0, 0,  \cdots0,h,-h}^{\mathrm{from} ~  1\mathrm{th} ~ \mathrm{to} ~ (y_{2}+1)\mathrm{th}},0,\cdots,0]^{T} .
\end{eqnarray}
Conducting the same matrix transformation as in (16), multiplying  (26) $\sim$ (29)  on the left-hand side by the known matrix  {\bf P}$_{m} $, we obtain
\begin{eqnarray}
{\bf X}_{n-1}+{\bf X}_{1}&=& diag(t_{1},t_{2},\cdots t_{m}){\bf X}_{0}-hJ{\bf D_{1}},\\
{\bf X}_{x-1}+{\bf X}_{x+1}&=& diag(t_{1},t_{2},\cdots t_{m}){\bf X}_{x}-hJ{\bf D_{2}},
\end{eqnarray}
with
\begin{eqnarray}
  {\bf D_{1}}&=&[\zeta_{1,1},\zeta_{1,2},\cdots,\zeta_{1,i},\cdots,\zeta_{1,m-1},\zeta_{1,m}]^{T},\nonumber\\
  \zeta_{1,i}&=&-\cos(y_{1}-\frac{1}{2})\theta_{i}+\cos(y_{1}+\frac{1}{2})\theta_{i} \nonumber\\
  &=& -2\sin(\frac{1}{2}\theta_{i})\sin(y_{1}\theta_{i}).
\end{eqnarray}
\begin{eqnarray}
  {\bf D_{2}}&=&[\zeta_{2,1},\zeta_{2,2},\cdots,\zeta_{2,i},\cdots,\zeta_{2,m-1},\zeta_{2,m}]^{T}, \nonumber\\
  \zeta_{2,i}&=&\cos(y_{2}-\frac{1}{2})\theta_{i}-\cos(y_{2}+\frac{1}{2})\theta_{i} \nonumber\\
  &=& 2\sin(\frac{1}{2}\theta_{i})\sin(y_{2}\theta_{i}).
\end{eqnarray}
Making use of the cyclicity, there must be $I_{n}^{(k)}=I_{0}^{(k)} $  $( k=1,2,\cdots, m)$. Now, let $k=n$  in (25) and obtain.
\begin{eqnarray}
X_{0}^{(i)}= X_{n}^{(i)}=X_{x+1}^{(i)}F_{n-x}^{(i)} -X_{x}^{(i)}F_{n-x-1}^{(i)}.
\end{eqnarray}

To obtain the initial conditions $X_{0}^{(i)}, X_{x}^{(i)}$  needed in our resistance calculation (11), we also need several independent equations. Since $X_{n-1}$  satisfies (25), and $X_{x-1}, X_{x}$  satisfy (24),  we therefore obtain three independent equations.  Together with (30), (31) and (34) we have six independent equations relating the six unknowns
$X_0^{(i)} ,X_1^{(i)} ,X_{n-1}^{(i)} ,X_{x-1}^{(i)} ,X_{x}^{(i)} ,X_{x + 1}^{(i)} $,   namely,
\begin{eqnarray}
\left( {\begin{array}{*{20}c}
   {\begin{array}{*{20}c}
   { - 1} & {t_i } & { - 1}  \\
   0 & {F_{x - 2}^{(i)} } & { - F_{x - 1}^{(i)} }  \\
   0 & {F_{x - 1}^{(i)} } & { - F_x^{(i)} }  \\
\end{array}} & {\begin{array}{*{20}c}
   0 & \qquad  0  & \qquad \quad 0  \\
   1 & \qquad  0  & \qquad \quad 0  \\
   0 & \qquad  1  & \qquad \quad 0  \\
\end{array}}  \\
   {\begin{array}{*{20}c}
   1 & \qquad 0 & \qquad 0  \\
   0 & \qquad 1 & \qquad 0  \\
   0 & \qquad 0 & \qquad 0  \\
\end{array}} & {\begin{array}{*{20}c}
   0 & {F_{n - x - 2}^{(i)} } & { - F_{n - x - 1}^{(i)} }  \\
   0 & {F_{n - x-1}^{(i)} } & { - F_{n - x}^{(i)} }  \\
   { - 1} & {t_i } & { - 1}  \\
\end{array}}  \\
\end{array}} \right)\left( \begin{array}{l}
 X_{n - 1}^{(i)}  \\
 X_0^{(i)}  \\
 X_1^{(i)}  \\
 X_{x - 1}^{(i)}  \\
 X_x^{(i)}  \\
 X_{x + 1}^{(i)}  \\
 \end{array} \right) = \left( \begin{array}{l}
 hJ\zeta _{1,i}  \\
   \quad 0 \\
   \quad 0 \\
   \quad 0 \\
   \quad 0 \\
 hJ\zeta _{2,i}  \\
 \end{array} \right)
\end{eqnarray}
where $t_i  = 2(1 + h) - 2h\cos \theta _i$  is given by (18). Solving (35) finally obtain after some algebra and reduction the two solutions needed in our resistance calculation (11),
\begin{eqnarray}
X_{0}^{(i)}&=& \frac{(F_{n-x}^{(i)}+F_{x}^{(i)})\zeta_{2,i}+F_{n}^{(i)}\zeta_{1,i}}
 {\lambda _i^n  + \bar \lambda _i^n  - 2}(hJ)  \nonumber \\
  &=&  2hJ\bigg[ \frac{(F_{n-x}^{(i)}+F_{x}^{(i)})\sin(y_2\theta_i)-F_{n}^{(i)}\sin(y_1\theta_i) }
 {\lambda _i^n  + \bar \lambda _i^n  - 2} \bigg]   \sin ( \frac{1}{2}\theta_i ) , \\
 X_{x}^{(i)}&=& \frac{(F_{n-x}^{(i)}+F_{x}^{(i)})\zeta_{1,i}+F_{n}^{(i)}\zeta_{2,i}}
                 {\lambda _i^n  + \bar \lambda _i^n  - 2}(hJ) \nonumber \\
           &=&  2hJ\bigg[ \frac{F_{n}^{(i)}\sin(y_2\theta_i)-(F_{n-x}^{(i)}+F_{x}^{(i)})\sin(y_1\theta_i) }
 {\lambda _i^n  + \bar \lambda _i^n  - 2} \bigg]   \sin (\frac{1}{2}\theta_i )  ,
\end{eqnarray}
Eq.(36) and (37) are two pivotal formulae needed in our resistance calculation (11).

\subsection{3.4 Derivation of the general formula }
From (11) it is clear that the currents ( $I_{0}^{(i)}, I_{x}^{(i)}$ ) must be calculated for evaluating  the equivalent resistance $R_{m\times n}(d_{1}, d_{2})$ . Making use of (20) and (21),  and conducting the matrix  inverse transformation  yields
\begin{eqnarray}
\left( {\begin{array}{cccc}
   {{I_{k}^{(1)}}} \\ {{I_{k}^{(2)}}} \\\vdots \\{{I_{k}^{(m)}}}\\
\end{array}}\right)=\frac{2}{m}
\left({\begin{array}{cccc}
   {\cos(1-\frac{1}{2})\theta_{1}} & {\cos(1-\frac{1}{2})\theta_{2}} &{\cdots} & {\cos(1-\frac{1}{2})\theta_{m}} \\
   {\cos(2-\frac{1}{2})\theta_{1}} & {\cos(2-\frac{1}{2})\theta_{2}} & {\cdots}  & {\cos(2-\frac{1}{2})\theta_{m}} \\
   {\vdots} & {\vdots} & {\ddots}   &{\vdots} \\
   {\cos(m-\frac{1}{2})\theta_{1}} & {\cos(m-\frac{1}{2})\theta_{2}} & {\cdots} & {\cos(m-\frac{1}{2})\theta_{m}} \\
\end{array}} \right)
\left( {\begin{array}{cccc}
   {{X_{k}^{(1)}}} \\ {{X_{k}^{(2)}}} \\\vdots \\{{X_{k}^{(m)}}} \\
\end{array}} \right) .
\end{eqnarray}
By (38) with $k=0 $  we achieve the following equation
\begin{equation}
\sum_{j=1}^{y_{1}}I_{0}^{(j)}=\frac{2}{m}\sum_{i=1}^{m}X_{0}^{(i)}\sum_{j=1}^{y_{1}}\cos(j-\frac{1}{2})\theta_{i})
=\frac{1}{m}\sum_{i=1}^{m}X_{0}^{(i)}\frac{\sin(y_{1}\theta_{i})}{\sin(\theta_{i}/2)}.
\end{equation}
where the following formula is used,
$$\sum\limits_{j = 1}^y {\cos (j- \frac{1}{2})\theta _i }  = {{\sin y\theta _i } \mathord{\left/
 {\vphantom {{\sin y\theta _i } {[2\sin (\frac{1}{2}\theta _i )]}}} \right.
 \kern-\nulldelimiterspace} {[2\sin (\frac{1}{2}\theta _i )]}} . $$
Similarly, we also obtain
\begin{eqnarray}
\sum_{j=1}^{y_{2}}I_{x}^{(j)}=\frac{1}{m}\sum_{i=1}^{m}X_{x}^{(i)}\frac{\sin(y_{2}\theta_{i})}{\sin(\theta_{i}/2)}.
\end{eqnarray}
Substituting (39) and (40) into (11),  we obtain
\begin{eqnarray}
R_{m\times n}(d_{1},d_{2})=\frac{r_{0}}{Jm}\sum_{i=1}^{m}\frac{X_{x}^{(i)}\sin(y_{2}\theta_{i})-X_{0}^{(i)}\sin(y_{1}\theta_{i})}{\sin(\theta_{i}/2)}.
\end{eqnarray}
Finally, we obtain the main result (2) by further substituting $X_{1}^{(i)}$  and $X_{x+1}^{(i)}$  from (36) and (37) into (41).

\subsection{3.5 Several special cases }

 When $y_{1}, y_{2}$  and  $x$  are special coordinates,  several interesting result can be derived.

 Case 1: when $n\rightarrow\infty $,  as $\lambda _i  > 1 > \bar \lambda _i  > 0 $ , thus
 \begin{eqnarray}
\mathop {\lim }\limits_{n \to \infty } \frac{{F_n^{(i)} }}{{\lambda _i^n  + \bar \lambda _i^n  - 2}} = \mathop {\lim }\limits_{n \to \infty } \frac{{\lambda _i^n  - \bar \lambda _i^n }}{{(\lambda _i  - \bar \lambda _i )(\lambda _i^n  + \bar \lambda _i^n  - 2)}} = \frac{1}{{\lambda _i  - \bar \lambda _i }}, \nonumber\\
\mathop {\lim }\limits_{n \to \infty } \frac{{F_x^{(i)}  + F_{n - x}^{(i)} }}{{\lambda _i^n  + \bar \lambda _i^n  - 2}} = \mathop {\lim }\limits_{n \to \infty } \frac{{\lambda _i^x  - \bar \lambda _i^x  + \lambda _i^{n - x}  - \bar \lambda _i^{n - x} }}{{(\lambda _i  - \bar \lambda _i )(\lambda _i^n  + \bar \lambda _i^n  - 2)}} = \frac{{\bar \lambda _i^x }}{{\lambda _i  - \bar \lambda _i }} .
\end{eqnarray}
 Substituting (42) into (2), we obtain (5) after using  $\lambda _i  - \bar \lambda _i  = 2\sqrt {(1 + h - h\cos \theta _i )^2  - 1} $.

 Case 2: If $\theta _k  = {(2k - 1)\pi }/ {(2m)}$, we have  $\Delta \theta _k  = \theta _{k + 1}  - \theta _k  = {\pi /m}$
. In the limit of $m\rightarrow \infty $ , we have
 \begin{eqnarray}
\mathop {\lim }\limits_{m \to \infty } \frac{1}{m}\sum\limits_{k = 1}^m {g(\theta _k )}  = \frac{1}{\pi }\int_0^\pi  {g(\theta )}  d\theta
\end{eqnarray}
which is an identity valid for any function  $g(\theta_{k})$. Eq.(43) is prepared to prove Eq.(6) in the following.

According to the identity transform of a trigonometric function, we have
\begin{eqnarray*}
S_{1,i}^2  + S_{2,i}^2  = 1 - \cos (y_1  + y_2 )\theta _i \cos (y_1  - y_2 )\theta _i , \nonumber\\
S_{1,i} S_{2,i}  = \frac{1}{2}[\cos (y_1  - y_2 )\theta _i  - \cos (y_2  + y_2 )\theta _i ] .
\end{eqnarray*}
When  $m\rightarrow \infty $ and $y_{1}, y_{2} \rightarrow \infty $ , we must transform
 $$\left( {y_1 ,y_2 } \right) = \left( {\frac{m}{2} + p, ~~ \frac{m}{2} + q} \right) , $$
where  $p,q \ll m$ are integers, and  $p-q=y_{1}-y_{2}$ are finite. As $\theta_{k}=(2k-1)\pi/(2m)$ ,  we have
$$\mathop {\lim }\limits_{m \to \infty } (y_1  + y_2 )\theta _i  = \mathop {\lim }\limits_{m \to \infty } (\frac{{m + p + q}}{{2m}})(2i - 1)\pi  = (i - \frac{1}{2})\pi . $$
Thus
\begin{eqnarray}
\mathop {\lim }\limits_{m \to \infty } (S_{1,i}^2  + S_{2,i}^2 ) = 1 - \sin (i\pi )\cos (y_1  - y_2 )\theta _i  = 1 ,\qquad \nonumber\\
\mathop {\lim }\limits_{m \to \infty } S_{1,i} S_{2,i}  = \frac{1}{2}[\cos (y_1  - y_2 )\theta _i  - \sin (i\pi )] = \frac{1}{2}\cos (y_1  - y_2 )\theta _i .
\end{eqnarray}
By (5) and (44) we therefore obtain
\begin{eqnarray}
R_{\infty  \times \infty }^{} \left( {\{ 0,y_1 \} ,\{ x,y_2 \} } \right) = \mathop {\lim }\limits_{m \to \infty } \frac{r}{m}\sum\limits_{i = 1}^m {\frac{{1 - \bar \lambda _i^x \cos (y_1  - y_2 )\theta _i }}{{\sqrt {(1 + h - h\cos \theta _i )^2  - 1} }}}  \nonumber\\
= \frac{r}{\pi }\int_0^\pi  {\frac{{1 - \bar \lambda _i^x \cos (y_1  - y_2 )\theta }}{{\sqrt {(1 + h - h\cos \theta )^2  - 1} }}} d\theta .\qquad
\end{eqnarray}
Thus Eq. (6) is proved.

 Case 3: When $d_{1} $ and $d_{2} $ are on the same radial, we have $x=0 $, then formula (2) reduces to
\begin{eqnarray}
R_{m \times n} (\{ 0,y_1 \} ,\{ 0,y_2 \} ) = \frac{{2r}}{m}\sum\limits_{i = 1}^m {\frac{{(\sin y_1 \theta _i  - \sin y_2 \theta _i )^2 F_n^{(i)} }}{{\lambda _i^n  + \bar \lambda _i^n  - 2}}} ,
\end{eqnarray}
where $F_{0}^{(i)}=0 $  is used. Since $F_{k}^{(i)}=(\lambda_{i}^{k}-\bar{\lambda}_{i}^{k})/(\lambda_{i}-\bar{\lambda}_{i})$, $\lambda_{i}\cdot\bar{\lambda}_{i}=1 $,  we have
\begin{eqnarray}
\frac{F_{n}^{(i)}} {\lambda_{i}^{n}+\bar{\lambda}_{i}^{n}-2}=\frac{\lambda_{i}^{n}-\bar{\lambda}_{i}^{n}}{(\lambda_{i}-\bar{\lambda}_{i})(\lambda_{i}^{n}+\bar{\lambda}_{i}^{n}-2)}\qquad\nonumber\\
=\frac{\lambda_{i}^{n/2}+\bar{\lambda}_{i}^{n/2}}{(\lambda_{i}-\bar{\lambda}_{i})(\lambda_{i}^{n/2}-\bar{\lambda}_{i}^{n/2})}=\frac{\coth(n\ln\sqrt{\lambda_{i}})}{\lambda_{i}-\bar{\lambda}_{i}}.
\end{eqnarray}
Substituting (47) into (46) , we immediately reduce to (7) from (46).

Case 4:  when the nodes  $d_{1} $ and $d_{2} $   are respectively at the center and the edge, we know $x=0 $, $y_{1}=0 $ and  $y_{2}=m $. As $\theta_{i}=(2k-1)\pi/(2m)$,  we therefore have
$$(\sin y_{1}\theta_{i}-\sin y_{2}\theta_{i})^{2}=(\sin 0 -\sin m\theta_{i})^{2}=1 $$
Obviously,  (7) immediately reduces to  (8).

Case 5: When  $d_{1}, d_{2}$  are on the same arc line, we have $y_{1}=y_{2} $. Thus  formula (2) immediately reduces to  (9).

Case 6: when the nodes   $d_{1} $ and $d_{2} $   are all at the edge, there be $y_{1}=y_{2}=m $, we therefore have
$$ (\sin y\theta_{i})^{2}=(\sin m\theta_{i})^{2}=1 . $$
Obviously, from (2) we immediately obtain (10).

\subsection{3.6 Two simple cases and testing }

{\bf Example 1. }
When $m=1$, the fig.1 degrades into a n-side polygon network as shown in fig.3. From [7] the resistance is given as
\begin{eqnarray}
R(O,A_{k})=r_{1}\frac{\coth(n\ln\sqrt\alpha)}{\alpha-\beta}.
\end{eqnarray}
where $A_{k}$ is  a arbitrary node on the boundary, and the boundary resistor is $r_{1}$, and $\alpha $ is given by  $(\alpha\cdot\beta=1 )$
\begin{eqnarray}
\alpha=1+\frac{1}{2}h_{1}+\sqrt{(1+\frac{1}{2}h_{1})^{2}-1}.
\end{eqnarray}
where $h_{1}=r_{1}/r_{0}$.  When $r_{1}=2r$, from (49) we have
\begin{eqnarray}
\lambda_{1}=\alpha=1+ h +\sqrt{(1+ h )^{2}-1}.\nonumber
\end{eqnarray}
where $h=r/r_{0}$ , and $\lambda_{1}$  is defined in (4). Obviously, (48)  is verified by (48) in the case of $m =1$.

\begin{figure*}
\begin{center}
\includegraphics[width=6cm,bb=0 0 146 140]{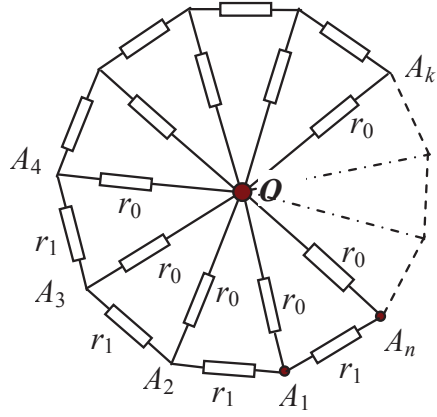}
\caption{ A $1\times n$ polygon network model, which has $n$ radial line and one polygon. Bonds in the radial and boundary are  respectively resistors $r_{0}$ and $r_{1}$ . }
\end{center}
\end{figure*}

{\bf Example 2}. we apply (2) to a $2\times 4$  cobweb network  with a $2r$ boundary shown in Fig.4. In this case the summation in (2) has two term $i=1,2$  with $\theta_{1}=\pi/4$  and $\theta_{2}=3\pi/4$ , and
\begin{eqnarray*}
F_{0}^{(i)}=0, \quad   F_{1}^{(i)}=1, \quad  F_{2}^{(i)}=2 + 2h - 2h\cos \theta _i , \nonumber\\
\lambda _i  + \bar \lambda _i  = 2 + 2h - 2h\cos \theta _i , \qquad  \qquad   \nonumber\\
\lambda _i  - \bar \lambda _i  = 2\sqrt {(1 + h - h\cos \theta _i )^2  - 1} , \qquad   \nonumber\\
\frac{{F_4^{(i)}  - F_3^{(i)}  - 1}}{{(\lambda _i^2  - \bar \lambda _i^2 )^2 }}
 = \frac{{\lambda _i^{}  + \bar \lambda _i^{}  + 1}}{{(\lambda _i^{}  + \bar \lambda _i^{} )(\lambda _i^{}  + \bar \lambda _i^{}  + 2)}}  \qquad \qquad   \nonumber\\
= \frac{{3 + 2h - 2h\cos \theta _i }}{{(2 + 2h - 2h\cos \theta _i )(4 + 2h - 2h\cos \theta _i )}}.
\end{eqnarray*}
As $\coth (n\ln \sqrt {\lambda _i } ) = (\lambda _i^{n/2}+ \bar \lambda _i^{n/2})/(\lambda _i^{n/2}- \bar \lambda _i^{n/2}) $, thus we have
\begin{eqnarray}
\frac{{\coth (4\ln \sqrt {\lambda _i } )}}{{\lambda _i  - \bar \lambda _i }} = \frac{{\lambda _i^2  + \bar \lambda _i^2 }}{{(\lambda _i  - \bar \lambda _i )(\lambda _i^2  - \bar \lambda _i^2 )}} = \frac{{(\lambda _i  + \bar \lambda _i )^2  - 2}}{{(\lambda _i  + \bar \lambda _i )(\lambda _i  - \bar \lambda _i )^2 }} \nonumber\\
 = \frac{{2(1 + h - h\cos \theta _k )^2  - 1}}{{4(1 + h - h\cos \theta _k )[(1 + h - h\cos \theta _k )^2  - 1]}}\nonumber\\
= \frac{{2(1 + h - h\cos \theta _k )^2  - 1}}{{4h(1 - \cos \theta _k )(1 + h - h\cos \theta _k )(2 + h - h\cos \theta _k )}} .
\end{eqnarray}

For the resistance between $O$ and $A_{1}$, we use (8) with $m=2$, and obtain
\begin{eqnarray}
&& R(O,A_1 ) = r\left( {\frac{{\coth (4\ln \sqrt {\lambda _1 } )}}{{\lambda _1  - \bar \lambda _1 }} + \frac{{\coth (4\ln \sqrt {\lambda _2 } )}}{{\lambda _2  - \bar \lambda _2 }}} \right) \nonumber\\
&& = \frac{{r_0 }}{4}\left( {\frac{{2(1 + h - h\cos \theta _1 )^2  - 1}}{{(1 - \cos \theta _1 )(1 + h - h\cos \theta _1 )(2 + h - h\cos \theta _1 )}}} \right.\nonumber\\
&& \qquad \left.{ + \frac{{2(1 + h - h\cos \theta _2 )^2  - 1}}{{(1 - \cos \theta _2 )(1 + h - h\cos \theta _2 )(2 + h - h\cos \theta _2 )}}}\right) \nonumber\\
&&  = 2r_0 \left( {\frac{{h^4  + 9h^3  + 21h^2  + 17h + 4}}{{(h^2  + 4h + 2)(h^2  + 8h + 8)}}} \right) ,
\end{eqnarray}
where $\theta_{1}=\pi/4$ and $\theta_{2}=3\pi/4$ are used.

\begin{figure*}
\begin{center}
\includegraphics[width=6cm,bb=0 0 204 193]{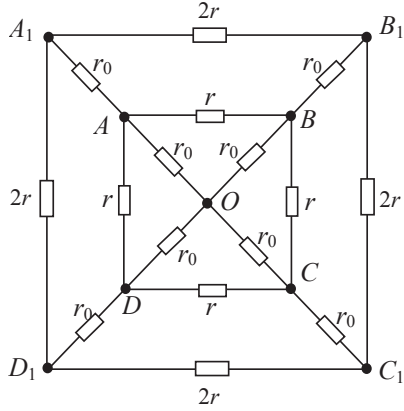}
\caption{  A $2\times 4$ polygon network, which has 4 radial line and two polygon. Bonds in the edge are resistor  $r_{1}=2r $. }
\end{center}
\end{figure*}

For the resistance between O and  $A$, we use (7) with $y_{1}=0, y_{2}=1 $ , and obtain
\begin{eqnarray}
R(O,A) = r\left( {\sin ^2 \theta _1 \frac{{\coth (4\ln \sqrt {\lambda _1 } )}}{{\lambda _1  - \bar \lambda _1 }} + \sin ^2 \theta _2 \frac{{\coth (4\ln \sqrt {\lambda _2 } )}}{{\lambda _2  - \bar \lambda _2 }}} \right) \nonumber\\
= \frac{r}{2}\left( {\frac{{\coth (4\ln \sqrt {\lambda _1 } )}}{{\lambda _1  - \bar \lambda _1 }} + \frac{{\coth (4\ln \sqrt {\lambda _2 } )}}{{\lambda _2  - \bar \lambda _2 }}} \right) \qquad  \qquad   \nonumber\\
= \frac{1}{2}R(O,A_1 ) = r_0 \left( {\frac{{h^4  + 9h^3  + 21h^2  + 17h + 4}}{{(h^2  + 4h + 2)(h^2  + 8h + 8)}}} \right).
\end{eqnarray}
where $R(O,A_{1})= 2 R(O,A)$  is discovered.

For the resistance between  $A$ and $B$ , we use (9) with $x=1, y=1 $, and obtain
\begin{eqnarray}
&&  \frac{{R_{m \times n} \left( {A,B} \right)}}{r} = 2\left( {\frac{{F_4^{(1)}  - (F_1^{(1)}  + F_3^{(1)} )}}{{(\lambda _1^2  - \bar \lambda _1^2 )^2 }}S_{1,1}^2  + \frac{{F_4^{(2)}  - (F_1^{(2)}  + F_3^{(2)} )}}{{(\lambda _2^2  - \bar \lambda _2^2 )^2 }}S_{1,2}^2 } \right)  \nonumber\\
&& \qquad  = \frac{{F_4^{(1)}  - (F_1^{(1)}  + F_3^{(1)} )}}{{(\lambda _1^2  - \bar \lambda _1^2 )^2 }} + \frac{{F_4^{(2)}  - (F_1^{(2)}  + F_3^{(2)} )}}{{(\lambda _2^2  - \bar \lambda _2^2 )^2 }}  \nonumber\\
&& \qquad  = \frac{{3 + 2h - 2h\cos \theta _1 }}{{(2 + 2h - 2h\cos \theta _1 )(4 + 2h - 2h\cos \theta _1 )}} \nonumber\\
&& \qquad  \qquad  + \frac{{3 + 2h - 2h\cos \theta _2 }}{{(2 + 2h - 2h\cos \theta _2 )(4 + 2h - 2h\cos \theta _2 )}} \nonumber\\
&&  \qquad  = \frac{{(3 + 2h)(h^2  + 6h + 4)}}{{(h^2  + 4h + 2)(h^2  + 8h + 8)}}
\end{eqnarray}
where $\theta_{1}=\pi/4$ and $\theta_{2}=3\pi/4$ are used.

For the resistance between $A_{1}$  and $B_{1}$, we use (10) with $x=1, y=2 $ , and obtain
\begin{eqnarray}
\frac{{R_{m \times n}^{{\rm{edge}}} (A_1 ,B_1 )}}{r} &=& 2\left( {\frac{{F_4^{(1)}  - (F_1^{(1)}  + F_3^{(1)} )}}{{(\lambda _1^2  - \bar \lambda _1^2 )^2 }} + \frac{{F_4^{(2)}  - (F_1^{(2)}  + F_3^{(2)} )}}{{(\lambda _2^2  - \bar \lambda _2^2 )^2 }}} \right) \nonumber\\
&=& 2\frac{{R_{mn}^{} \left( {A,B} \right)}}{r} = \frac{{2(3 + 2h)(h^2  + 6h + 4)}}{{(h^2  + 4h + 2)(h^2  + 8h + 8)}} .
\end{eqnarray}
where $R(A_{1},B_{1})= 2 R(A,B)$  is discovered.

For the resistance between  $A_{1}$ and $C_{1}$ , we use (10) with  $x=2, y=2 $ , and obtain
\begin{eqnarray}
\frac{{R_{m \times n}^{{\rm{edge}}} (A_1 ,C_1 )}}{r} &=&  2\left( {\frac{{F_4^{(1)}  - 2F_2^{(1)} }}{{(\lambda _1^2  - \bar \lambda _1^2 )^2 }} + \frac{{F_4^{(2)}  - 2F_2^{(2)} }}{{(\lambda _2^2  - \bar \lambda _2^2 )^2 }}} \right) \nonumber\\
&=&  2\left( {\frac{1}{{\lambda _1  + \bar \lambda _1 }} + \frac{1}{{\lambda _2  + \bar \lambda _2 }}} \right) \nonumber\\
 &=&  2\left( {\frac{1}{{2 + 2h - 2h\cos \theta _1 }} + \frac{1}{{2 + 2h - 2h\cos \theta _2 }}} \right)  \nonumber\\
&=&  = \frac{{4(1 + h)}}{{h^2  + 4h + 2}} .
\end{eqnarray}

For the resistance between $A$  and $C$, we use (10) with   $x=2, y=1$ , and obtain
\begin{eqnarray}
\frac{{R_{m \times n}^{} (A,C)}}{r} = \frac{{F_4^{(1)}  - 2F_2^{(1)} }}{{(\lambda _1^2  - \bar \lambda _1^2 )^2 }} + \frac{{F_4^{(2)}  - 2F_2^{(2)} }}{{(\lambda _2^2  - \bar \lambda _2^2 )^2 }} \nonumber\\
 = \frac{1}{2}\frac{{R_{m \times n}^{{\rm{edge}}} (A_1 ,C_1 )}}{r} = \frac{{2(1 + h)}}{{h^2  + 4h + 2}} ,
\end{eqnarray}
where $R(A_{1}, C_{1})= 2 R(A, C)$  is discovered.
Here $ A_{1},A, B_{1},B, C_{1},C, D_{1},D $  denote nodes shown in Fig.4 and we have used $h=r/r_{0}$. We have verified these results by carrying out explicit calculations in the actual circuit.

\subsection{3.7  Expressing (2) by hyperbolic function }
We find that (2) can be expressed by the hyperbolic functions. Defining
\begin{eqnarray}
\lambda _i = e^{2L_i }  = 1 + h - h\cos \theta _i  + \sqrt {(1 + h - h\cos \theta _i )^2  - 1} , \nonumber\\
\bar \lambda _i  = e^{ - 2L_i }  = 1 + h - h\cos \theta _i  - \sqrt {(1 + h - h\cos \theta _i )^2  - 1} , \nonumber\\
\cosh (2L_i ) = 1 + h - h\cos \theta _i .\qquad \qquad \qquad
\end{eqnarray}
Then there be
\begin{eqnarray}
\sinh (2kL_i ) = \frac{1}{2}(\lambda _i^k  - \bar \lambda _i^k ),\quad
\cosh (2kL_i ) = \frac{1}{2}(\lambda _i^k  + \bar \lambda _i^k ),
\end{eqnarray}
and
\begin{eqnarray}
F_n^{(i)}  = \frac{{\lambda _i^k  - \bar \lambda _i^k }}{{\lambda _i^{}  - \bar \lambda _i^{} }} &=& \frac{{\sinh 2nL_i }}{{\sinh 2L_i }} = \frac{{2\sinh( nL_i) \cosh (nL_i )}}{{\sinh 2L_i }}, \nonumber\\
F_x^{(i)}  + F_{n - x}^{(i)} &=&  \frac{{2\sinh (nL_i )\cosh [(n - 2x)L_i] }}{{\sinh 2L_i }}, \nonumber\\
\lambda _i^n  + \bar \lambda _i^n  - 2 &=& (\lambda _i^{n/2}  - \bar \lambda _i^{n/2} )^2  = 4\sinh ^2 (nL_i ).
\end{eqnarray}
Therefore,  simplifying (2) to reduce to
\begin{eqnarray}
R_{m\times n} (d_1,  d_2 ) =
 \frac{{r}}{m}\sum\limits_{i = 1}^m {\frac{{\cosh (nL_i )( S_{1,i}^{2} + S_{2,i}^{2} ) - 2\cosh [(n - 2x)L_i ]S_{1,i} S_{2,i} }}{{\sinh 2L_i \sinh nL_i }}} .
\end{eqnarray}
where $S_{k,i}= \sin (y_{k}\theta_{i})$. Formula (60) is equivalent to the formula (2).

\section{4.  Summary and discussion}

The computation of the two-point resistance of an $m\times n$ resistor network has always been a problem even though the problem has been studied more than 160 years. In 2004 Wu [3] established a theorem to compute the equivalent resistance of an $m\times n$ resistor network. Using the theorem many results have been obtained, but the results are always in the form of a double summation. The additional work required to reduce this to a single summation can be quite complex. Besides, the Wu's method is difficult to solve the resistor network with different boundary such as the problem of this paper [12].

An alternative direct approach of computing resistances had been developed by us [7-9] which, when applied to the cobweb and rectangular networks, gives the results in terms of a single summation, thus offering a direct and somewhat simpler approach. The direct method has been used  by several authors  to deduce the two-point resistance in a fan and a globe networks [11,12]. Here we use the direct method to compute resistances in a cobweb network with a $2r$ boundary..

It is necessary for us to consider a profound question how to calculate the equivalent resistance of the resistor cobweb network with arbitrary boundary, this problem is equivalent to such a problem that how to find explicit eigenvector of matrix $A_{m}$  in (1) when $h_{1}$  is an arbitrary constant. We are looking forward to the final solution of the problem in the future.

\section*{Acknowledgment}
This work is supported by Jiangsu Province Education Science Plan Project (No. D/2013/01/048), the Research Project for Higher Education Research of Nantong University (No. 2012GJ003).

\end{document}